\documentclass[12pt]{article}
\usepackage{graphicx}
\def\np{Nucl.\ Phys.\ }

\def\prl{Phys.\ Rev.\ Lett.\ }
\def\cmp{Commun.\ Math.\ Phys.\ }
\def\prb{Phys.\ Rev.\ B }
\def\neel{N\'{e}el }
\begin{document}
\newcommand{\tb}{\tilde{b}}
\newcommand{\tm}{\tilde{m}}
\newcommand{\br}{{\bf r}}
\newcommand{\bR}{{\bf R}}
\newcommand{\bk}{{\bf k}}
\newcommand{\cj}{{\cal J}}
\newcommand{\half}{\mbox{$\frac{1}{2}$}}
\newcommand{\etp}{\eta_{p}}
\newcommand{\mf}{\mbox{ (mod 4) }}
\newcommand{\mt}{\mbox{ (mod 2) }}
\newcommand{\bqo}{\bar{Q}_1}
\newcommand{\blam}{\bar{\lambda}}
\newcommand{\ttau}{\tilde{\tau}}
\newcommand{\hx}{\hat{x}}
\newcommand{\hy}{\hat{y}}
\setlength{\baselineskip}{.375in}
\vspace{1in}
\begin{center}
\Large\bf
Large $N$ expansion for frustrated and doped quantum
antiferromagnets\\
\vspace{0.75in}
\normalsize\rm Subir Sachdev and N. Read\\
\vspace{0.325in}
\normalsize\em
Center for Theoretical Physics, P.O. Box 6666\\
and\\
Department of Applied Physics, P.O. Box 2157\\
Yale University, New Haven, CT 06511\\
\vspace{0.55in}
\end{center}
\rm
A large $N$ expansion technique, based on symplectic
($Sp(N)$) symmetry, for
frustrated magnetic systems is studied. The phase diagram
of a
square lattice, spin $S$,
quantum antiferromagnet with first, second and third
neighbor antiferromagnetic
coupling (the $J_1$-$J_2$-$J_3$ model)
is determined in the large-$N$
limit and
consequences of fluctuations at finite $N$ for the quantum
disordered phases are discussed. In addition to phases with long range
magnetic order, two classes of disordered phases
are found: {\em (i)} states similar to those in unfrustrated
systems with commensurate, collinear spin correlations,
confinement of spinons, and
spin-Peierls or valence-bond-solid order controlled by
the value of $2S \mf$ or $2S \mt$; {\em (ii)} states with
incommensurate, coplanar spin correlations, and unconfined
bosonic spin-$1/2$ spinon excitations. The occurrence of
``order from disorder'' at large $S$ is discussed.
Neither chirally ordered
nor spin nematic
states are found. Initial results on superconductivity in
the
$t-J$ model at $N=\infty$ and zero temperature are also
presented.
\begin{verbatim}
PACS Nos. 75.10.Jm, 75.50.Ee, 74.65.+n
September, 1990
International Journal of Modern Physics B 5, 219, (1991).
\end{verbatim}
\newpage
\section*{\Large\bf 1. Introduction}

Spurred by the discovery of high temperature
superconductivity~\cite{nobel}, the
last few years have seen intense activity in the subject of
two-dimensional quantum antiferromagnets. In particular,
much attention
has been focussed upon understanding the structure of
quantum
disordered phases of such
antiferromagnets~\cite{sudip} and their
possible
relationship to
high temperature
superconductivity~\cite{anderson}.
Large $N$ expansions~\cite{assa}
on unfrustrated antiferromagnets
found columnar spin-Peierls order in the disordered
state for half-integer spins~\cite{hedge,self1,self2}.
``Chiral''~\cite{wwz} and ``spin-nematic''~\cite{premi}
states are among the new structures that have
been proposed as ground states of frustrated
quantum antiferromagnets.

This paper presents details of a new systematic analytic
technique for
frustrated and
doped antiferromagnets (AFMs) which has been proposed
recently~\cite{self3}.
The method relies on a large $N$ expansion based upon models
with
symplectic symmetry ($Sp(N)$).
The method is applied in this paper to two models:
\newline
({\it i\/}) A square lattice AFM with first,
second and third neighbor antiferromagnetic coupling - the
$J_1$-$J_2$-$J_3$ model. This model has been studied
elsewhere
by
numerical~\cite{elbio,numj3},
series-expansion~\cite{rajiv}, and
mean-field~\cite{rnb} methods; we will compare our results
with
these
studies later.
A particular strength of our approach is that it makes the
connection
between the structure of the known classically ordered
phases and
the quantum disordered phases especially clear. It will also
clarify
the appearance of
``order-from-disorder''~\cite{henley} from quantum
fluctuations
for models which are classically disordered.
We will find new quantum disordered phases which have
unconfined bosonic spin-1/2 spinon excitations.
\newline
({\it ii\/}) The square lattice $t-J$ model. We will present
initial
results in the mean field $N=\infty$ limit at $T=0$.
We find superconducting ground states similar in structure
to those
of Ref~\cite{bza}.

We begin by motivating the use of a symplectic large $N$
expansion
for frustrated AFMs.
Previous
large $N$ methods have been based on the $SU(N)$
generalization
of the $SU(2)$ AFM~\cite{assa,self1,self2} and have been
restricted
to
unfrustrated AFMs with a two sublattice structure (labeled
$A$,$B$)
for reasons we now explore.
``Spins'' are placed on sublattice $A$ forming an
irreducible
representation
of $SU(N)$ while those on $B$ form the conjugate
representation. Thus {\em e.g.} totally symmetric
representations
on sublattice $A$ can be
can be formed~\cite{assa}
by placing $n_b$ bosons
created by
$b^\dagger_{i\alpha}$ where $\alpha=1\ldots N$ and $i\in
A$; the
$b^{\alpha}$ bosons transform under the fundamental
representation of
$SU(N)$. The conjugate representation is placed on
sublattice $B$
by $n_b$ bosons
created by
$\bar{b}^{\dagger\alpha}_j$ for $j\in B$; the
$\bar{b}_{\alpha}$ bosons transform under the representation
conjugate
to the fundamental.
The use of conjugate representations ensures a natural
pairing
between directions in spin space on the two sublattices.
The only possible coupling between sites
on opposite sublattices which is invariant under $SU(N)$
and
bilinear in the spin operators is
\begin{equation}
-(b^\dagger_{i\alpha}
\bar{b}^{\dagger\alpha}_j)
(b^\beta_i \bar{b}_{j\beta})
\end{equation}
(For $SU(2)$ this reduces to the usual  $S_i\cdot
S_j$
plus a constant).
In the classical limit ($n_b \rightarrow \infty$) this
coupling
will induce a \neel ground state
\begin{equation}
\prod_{i\in A} \left( b_{i1}^{\dagger} \right)^{n_b}
\prod_{j \in B} \left(\bar{b}_j^{1\dagger}\right)^{n_b} | 0
\rangle.
\end{equation}
In a system with strong quantum fluctuations, this coupling
tries to
maximize the number of valence bonds
\begin{equation}
\left( b_{i\alpha}^{\dagger} \bar{b}_j^{\dagger\alpha}
\right)
|0 \rangle
\end{equation}
between pairs of sites on opposite sublattices.
The only bilinear coupling between sites on the same
sublattice
is the `ferromagnetic' coupling
\begin{equation}
-(b^\dagger_{i\alpha} b^\alpha_{i'})(b^\dagger_{i'\beta}
b^\beta_i)
\label{ferrocp}
\end{equation}
which demands that the spins on sites $i$ and $i^\prime$
point
in the same direction. Reversing the sign of this term will
favor
states in which the spins point in {\em any two} unequal
directions; this is quite different from the
antiferromagnetic coupling which explicitly pairs directions
in spin space. Interactions between sites on the same and
opposite lattices are thus inequivalent and there is no
natural way of introducing a democratic antiferromagnetic
coupling between any two sites. Only for $SU(2)$ do the
ferromagnetic and antiferromagnetic couplings above become
equivalent.

A proper description of a frustrated AFM therefore requires
that all the spins transform under the same representation
of a group, and that two spins can combine to form a
singlet. For $SU(2)$ we can write $\bar{b}_{j\alpha} \equiv
\varepsilon_{\alpha\beta} b_j^{\beta}$
($\varepsilon^{\sigma\sigma^\prime}=-
\varepsilon^{\sigma^\prime\sigma}$
and $\varepsilon^{\uparrow\downarrow}=1$); the boson
annihilation operators on all
the sites now have an upper index. Singlet bonds
$\varepsilon^{\sigma\sigma^\prime}
b^\dagger_{i\sigma} b^\dagger_{j\sigma^\prime}$ can now be
formed between any two sites.
This structure has a natural generalization to the
symplectic groups $Sp(N)$ for all $N$.
These are the groups
of $2N\times2N$ unitary matrices $U$
such that
\begin{equation}
U^{T} \cj U = \cj
\label{dsp2}
\end{equation}
where
\begin{equation}
\cj_{\alpha\beta} = \cj^{\alpha\beta} = \left( \begin{array}{cccccc}
  & 1 & & & & \\
-1 &  & & & & \\
 & &  & 1  & & \\
 & & -1 &  & & \\
 & &  & & \ddots & \\
 & & & & & \ddots
\end{array} \right)
\end{equation}
is the generalization of the $\varepsilon$ tensor (note
$Sp(1) \cong SU(2)$). This
tensor
can be used to raise or lower indices on other
tensors; all representations are therefore self-conjugate.
``Spins'' can be created on each site by $n_b$ bosons
$b_{i\alpha}^{\dagger}$ where $\alpha = 1 \ldots 2N$.
The $b_{i}^{\alpha}$ bosons transform as the fundamental
representation of $Sp(N)$; the ``spins'' on the lattice
therefore belong to the symmetric product of $n_b$
fundamentals, which is also an irreducible representation.
Valence bonds
\begin{equation}
\cj^{\alpha\beta} b_{i\alpha}^{\dagger}
b_{j\alpha}^{\dagger}
\end{equation}
can be formed between any two sites; this operator is a
singlet under $Sp(N)$ because of (\ref{dsp2}).
An antiferromagnetic coupling should maximize the number of
such bonds between two sites; we will therefore consider
models described by the following Hamiltonian
\begin{equation}
H_{AF} = -\sum_{i>j} \frac{J_{ij}}{N} \left(
\cj^{\alpha\beta}
b_{i\alpha}^{\dagger} b_{j,\beta}^{\dagger} \right)
\left( \cj_{\gamma\delta} b_{i}^{\gamma} b_{j}^{\delta}
\right)
\label{hex}
\end{equation}
where $i,j$ run over the sites of an arbitrary lattice, and
$J_{ij}$ are antiferromagnetic exchange constants. We recall
the constraint
\begin{equation}
b_{i\alpha}^{\dagger} b_i^{\alpha} = n_b
\end{equation}
which must be imposed at every site.
For the group $Sp(1)$ this generates states with spin $S =
n_b /2$
at every site.

\begin{figure}
\centerline{\includegraphics[width=7.5in]{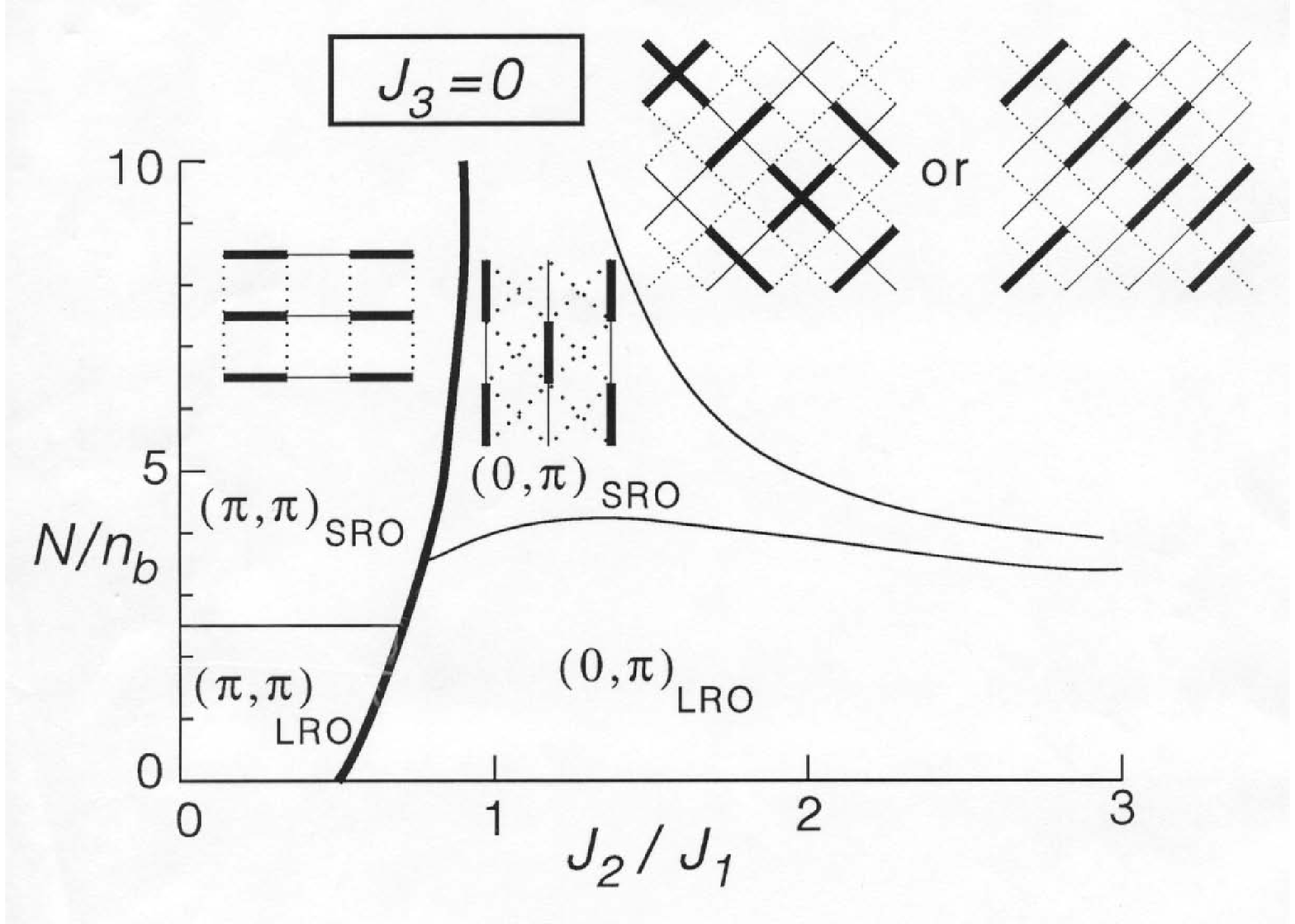}} \caption{
Ground states of $H$ for $J_3=0$ as a function of $J_2 / J_1$ and
$N/n_b$ ($n_b = 2S$ for $Sp(1) \equiv SU(2)$). Thick (thin) lines
denote first (second) order transitions at $N=\infty$. Phases are
identified by the wavevectors at which they have magnetic
long-range-order (LRO) or short-range-order (SRO). The links with
$Q_p\neq 0$ in each SRO phase are shown. The large $N/n_b$, large
$J_2 / J_1$ phase has the two sublattices decoupled at $N=\infty$;
each sublattice has \neel-type SRO. Spin-Peierls order at finite
$N$ for odd $n_b$ is illustrated by the thick, thin and dotted
lines. The $(\pi , \pi)$-SRO and the ``decoupled'' states have
line-type~\cite{self2} spin-Peierls order for $n_b =2 \mbox{(mod
4)}$ and are VBS for $n_b=0 \mbox{(mod 4)}$. The $(0,\pi )$-SRO
state is a VBS for all even $n_b$. } \label{fig1}
\end{figure}

We now summarize the results obtained in our study of
$H_{AF}$
for the model with nearest ($J_1$), second ($J_2$) and third
($J_3$) neighbor interactions on the square lattice. The
large $N$ limit was taken with the ratio $n_b /N$ fixed; the
results depend strongly on the value of $n_b /N$. The phases
found (Figs~\ref{fig1}, \ref{fig2}, \ref{fig3}, \ref{fig4})
can be separated into two distinct types:
\subsubsection*{\bf 1. Commensurate, collinear phases}
These are closely related to those found in unfrustrated
$SU(N)$ AFMs~\cite{self1,self2}. For large values of
$n_b / N$ these states have magnetic long range order (LRO)
with the spins polarized parallel or anti-parallel to each
other. Upon reducing $n_b / N$ a transition eventually
occurs
to a corresponding short range ordered
(SRO) phase, with a finite spin-correlation
length. All the SRO phases are
described at long distances and long times
by the following effective action when the
correlation length is not too small:
\begin{equation}
S_{eff} =
\int d^2 r \int_{0}^{c\beta}
d\ttau \left\{ \frac{1}{g} \left[
|(\partial_{\mu} - iA_{\mu})z^{\alpha}|^2
+ \frac{\Delta^2}{c^2}
|z^{\alpha} |^2\right]\right\} + \cdots,
\label{sefpp}
\end{equation}
where $\tau$ is the Matsubara time, $c$ is the spin-wave
velocity, $\ttau = c\tau$, and $\mu$ runs over $x,y,\ttau$.
Spatial anisotropy can be also be present
but has been
neglected for simplicity.
The action describes a complex field $z^{\alpha}$ which
transforms under the fundamental of $Sp(N)$ and has
a $U(1)$ charge $+1$; this field is related to the lattice
boson $b^{\alpha}$, although the form of the relationship
varies in different phases. The $U(1)$
gauge field $A_{\mu}$ is
related to the phases of certain link-variables, and the
gauge symmetry is compact. In the SRO phase, the
mass $\Delta$ is finite and
is proportional
to the inverse spin-correlation length.
The compact $U(1)$ gauge field leads to confinement of the
$z$ quanta and Berry phases of its
instantons (monopoles)~\cite{hedge} induce
spin-Peierls order for special values
of $n_b$~\cite{self2}.
The LRO phase is reached when $\Delta$ vanishes and
$z$ quanta condense in the $\bk_0 = 0$
state~\cite{yoshi}. To characterize the LRO phases by a
gauge-invariant order parameter, we consider the
expectation value of the `spin' operators
$s_a = \left\langle z_{\alpha}^{\ast} S_{a\beta}^{\alpha}
z^{\beta}\right\rangle $ ($S_a$ is a generator of $Sp(N)$)
across the transition.
In the SRO phase these vanish as a result
of $Sp(N)$
invariance. In the LRO phase, the order parameter manifold,
$M_{\rm coll}$  will be given by $Sp(N)$ modulo the subgroup
which leave the $s_a$ invariant.
Any orientation of the $z$ condensate
can be rotated by $Sp(N)$ transformations such that
$\langle z^{\alpha} \rangle = \bar{z}
\delta^{\alpha 1}$.
Now the $s_a$ are invariant under the group $
Sp(N-1)$ acting on components
$\alpha > 2$.
An additional factor of $U(1)$ is obtained
from the symmetry associated with rotations about the `$z$'
axis in the $\alpha = 1,2$ subspace under which
$z^1 \rightarrow z^1 e^{i\phi}$ and
$z^2 \rightarrow z^2 e^{-i\phi}$.
We have therefore
\begin{equation}
M_{\rm coll}={Sp(N)\over U(1)\times Sp(N-1)}
\end{equation}
in direct analogy with $U(N)/(U(1)\times U(N-1))\cong
CP^{N-1}$ for the $SU(N)$
models~\cite{self1,ian1};
for $N=1$ we recover $CP^{\,1}$ (the unit sphere, $S^2$).
The transition from LRO to SRO is expected to be described
by a non-linear sigma ($NL\sigma$) model on $M_{\rm coll}$.
We
note that
$\pi_2(M_{\rm coll})=Z$, the group of integers, so that
topologically stable
point defects in spacetime (hedgehogs) exist for all $N$
when the spatial
dimension $d=2$, while $\pi_1=0$ so there are no line
defects.
\subsubsection*{\bf 2. Incommensurate phases}
Frustration induces phases with LRO and SRO in
incommensurate
helical spin configurations. All of the phases found favor
planar arrangement of spins and there is no ``chiral''
order. In regions not too far from a commensurate phase
these states are described at long distances and long times
by the following effective
action:
\begin{equation}
S_{eff} =
\int d^2 r \int_{0}^{c\beta}
d\ttau \left\{ \frac{1}{g} \left[
|(\partial_{\mu} - iA_{\mu})z^{\alpha}|^2
+ \frac{\Delta^2}{c^2}
|z^{\alpha} |^2 \right] + \vec{\Phi} \cdot \left(
\cj_{\alpha\beta}z^{\alpha}\vec{\nabla} z^{\beta}\right) +
\mbox{c.c.}
+ V(\vec{\Phi}) \right\} + \dots,
\label{hgss}
\end{equation}
The new feature is the presence of a charge $-2$
two-component scalar
$\vec{\Phi} = (\Phi_x , \Phi_y)$ related to
lattice link fields which induce incommensurate order.
The coupling between the $z$
and $\vec{\Phi}$ is the simplest one consistent with
global $Sp(N)$ symmetry and is $U(1)$ gauge invariant due to the antisymmetry
of $\cj$ (a coupling with no
gradients of the form
$\vec{\Phi} \cj_{\alpha\beta} z^{\alpha} z^{\beta}$ vanishes
identically). The potential $V(\vec{ \Phi} )$
is induced by short wavelength
fluctuations; in the incommensurate phases
$V(\vec{ \Phi} )$ will have minima at a
non-zero value of $\vec{\Phi}$ leading to a Higgs phase
with
$\left\langle\vec{\Phi}\right\rangle \neq 0$.
All of phases found in this paper have at least two
gauge inequivalent minima related to each other by
a square lattice symmetry; this leads to a two-fold
degeneracy which can only be lifted by explicitly
breaking the square lattice symmetry in $H_{AF}$.
The minima are also such that
$\langle \Phi_x \rangle$
and $\langle \Phi_y \rangle$ can simultaneously be made
real in a suitable gauge. In such a gauge
we
see from Eqn (\ref{hgss}) that the
minima of the dispersion
of the $z^{\alpha}$ quanta are at wavevectors $\bk_0 = \pm
g(\langle\Phi_x \rangle
, \langle\Phi_y\rangle )/2$. For finite
$\Delta$ this leads to incommensurate SRO.
The $z^{\alpha}$ quanta have unit charge and will therefore
be unconfined~\cite{fradkin_shenker} in this Higgs phase:
these SRO phases therefore possess unconfined bosonic
spinons which transform under the fundamental of
$Sp(N)$~\cite{kol} for all values of the on-site
`spin' $n_b$.
In models which have the two-fold degeneracy lifted
by small explicit symmetry breaking terms in $H_{AF}$
these results appear to
contradict the `fractional quantization principle'
of Laughlin~\cite{laugh}.
As $\Delta \rightarrow 0$, the $z$ quanta
will condense at $\bk_0$ leading to incommensurate
LRO.
The order-parameter manifold is now
\begin{equation}
M_{\rm noncoll}=\frac{Sp(N)}{ Z_2\times Sp(N-1)}
\end{equation}
because the $U(1)$ invariance about the `$z$' axis has now
been reduced to $Z_2$ as the spins are no longer collinear.
This generalizes the result for $N=1$, $SO(3)\cong
SU(2)/Z_2$ pointed out
previously~\cite{halpsas,dombread}. For $M_{\rm noncoll}$,
$\pi_2=0$ but
$\pi_1=Z_2$, so there are line
defects in spacetime (vortex worldlines) for $d=2$ in this
case.
Instantons and vortices are suppressed in the Higgs phase,
so their Berry phases are not expected to lead to
spin-Peierls order. However the Berry phases could induce
additional intermediate phases between the Higgs and
confinement phases: this will be briefly discussed later.

The symplectic groups also have interesting
applications to doped AFMs as described {\em e.g.}
by a $t-J$ model.
In such a model, the hopping term
transfers spin from site to site and so resembles the
ferromagnetic coupling (\ref{ferrocp}).
Thus in $SU(N)$ models with
conjugate
representations on neighboring sites there is
no simple $SU(N)$-invariant hopping
term
for $N>2$; if instead one choses all sites
with the same
representation, hopping can be
included but not
antiferromagnetic exchange.
Our symplectic
approach allows inclusion of both exchange and hopping for
all $N$. In particular, using fermions with $Sp(N)$ indices
for spins and
bosons for holes, the large $N$ limit justifies the
decoupling of
Ref~\cite{bza} and produces superconductivity; phase
separation into an
insulating AFM and hole-rich
superconductor is also present. Details of this
analysis at $N=\infty$ and zero temperature
are presented in Section 4.

The outline of the rest of this paper is as follows.
In Section 2 we introduce the general formalism of
the $Sp(N)$ large $N$ limit for frustrated AFMs.
Section 3 presents detailed
results on the $J_1$-$J_2$-$J_3$ model. Finally
Section 4 discusses initial results on the $t$-$J$
model.

\section*{\Large\bf 2. Frustrated Antiferromagnets}

We begin by setting up the general framework for the large-$N$
expansion of antiferromagnets
with Hamiltonian $H_{AF}$ (Eqn (\ref{hex})).
The large-$N$ limit is taken with
$n_b /N$ fixed to an arbitrary value. Depending upon the
values of
the $J_{ij}$ and of $n_b / N$, the ground state of $H_{AF}$
may
either break global $Sp(N)$ symmetry and possess magnetic
long-range-order (LRO) or be $Sp(N)$ invariant with only
short-range magnetic order (SRO). The structure of the
large-$N$
limit is very similar to the $SU(N)$ case, and has been
discussed
extensively
before~\cite{assa,self2} for the SRO states. We will therefore
present details mainly in the LRO phases. The large $N$
limit
for these states is most conveniently taken by adapting the
method of Brezin and Zinn-Justin~\cite{blz}.
We begin by introducing the
parametrization
\begin{equation}
b_i^{m\sigma} \equiv \left( \begin{array}{c}
\sqrt{N} x_i^{\sigma} \\
\tb_i^{\tm\sigma} \end{array} \right)
\label{x}
\end{equation}
We have introduced a natural double-index notation
$\alpha \equiv (m, \sigma)$ with $m = 1 \ldots N$
and $\sigma = \uparrow , \downarrow$. The index
$\tm = 2, \ldots N$. The $x^{\sigma}$ field has been
introduced
to allow for a non-zero condensate
$\left\langle b_i^{m\sigma}\right\rangle = \sqrt{N}
\delta^m_1 x_i^{\sigma}$; we will only consider models in
which the condensate in the LRO phase can be transformed
by a uniform global $Sp (N)$ rotation into
this form.
We insert (\ref{x})
into $H_{AF}$, decouple the quartic
terms by
 Hubbard-Stratanovich fields $Q_{ij}$, and enforce the
constraints
by the Largrange multipliers $\lambda_i$. This yields
\begin{displaymath}
H_{MF} = \sum_{i>j} \left(
N J_{ij} \left| Q_{ij} \right|^2
- J_{ij} Q_{ij} \varepsilon_{\sigma\sigma^{\prime}}
\left( N x_i^{\sigma} x_{j}^{\sigma\prime} + \sum_{\tm}
\tb_{i}^{\tm\sigma}
\tb_{j}^{\tm\sigma^{\prime}} \right) + \mbox{H. c.} \right)
\end{displaymath}
\begin{equation}
~~~~~~~~~~~~~~~~ + \sum_{i} \lambda_i \left(
N |x_i^{\sigma}|^2 + \sum_{\tm} \tb_{i,\tm\sigma}^{\dagger}
\tb_{i}^{\tm\sigma} - n_b \right)
\end{equation}
The large $N$ limit is obtained by integrating over the
$2(N-1)$ $\tb$ fields.
The resulting effective action, expressed in terms of
the $Q_{ij}$, $\lambda_i$ and $x_i^{\sigma}$ fields,
will have a prefactor of $N$ (and some terms of order 1
which
are sub-dominant)
and is therefore well approximated by its saddle-point
value.
The $Q, \lambda, x$ fields are expected to be
time-independent at
the saddle-point and this is implicitly assumed in the
following.
The functional integral over the $\tb$ requires knowledge of
the
eigenmodes of $H_{MF}$. This can be done along standard
lines: we
first solve the eigenvalue equation
\begin{eqnarray}
\lambda_i U_{i\mu} - \sum_{j} J_{ij} Q_{ij}^{\ast} V_{j\mu}
& = &
\omega_{\mu}
U_{i\mu} \nonumber \\
\sum_{i} J_{ij} Q_{ij} U_{i\mu} - \lambda_j V_{j\mu} & = &
\omega_{\mu}
V_{j\mu}
\label{eig}
\end{eqnarray}
for the $N_s$ ($=$ number of sites in the system)
positive eigenvalues $\omega_{\mu}$ and the corresponding
eigenvectors
$(U_{i\mu}, V_{j\mu})$.
The bosonic eigenoperators $\gamma_{\mu
}^{\tm\sigma}$
\begin{equation}
\gamma_{\mu}^{\tm\sigma} = \sum_{i} \left( U_{i\mu}^{\ast}
\tb_{i}^{\tm\sigma} - \delta^{\tm\tm^{\prime}}
\varepsilon^{\sigma\sigma^{\prime}} V_{i\mu}^{\ast}
\tb_{i\tm^\prime \sigma^\prime}^{\dagger} \right)
\end{equation}
will diagonalize $H_{MF}$. The inverse relation is
\begin{equation}
\tb_{i}^{\tm\sigma} = \sum_{\mu} \left( U_{i\mu}
\gamma_{\mu}^{\tm\sigma} - \delta^{\tm\tm^{\prime}}
\varepsilon^{\sigma\sigma^{\prime}} V_{i\mu}^{\ast}
\gamma_{\mu\tm^\prime \sigma^\prime}^{\dagger} \right)
\end{equation}
Consistency of these relations imposes certain orthogonality
requirements on $(U , V)$. It can be shown using Eqn
(\ref{eig})
that these can always be satisfied.
Finally the ground state energy, $E_{MF}$ of $H_{MF}$
is shown to be
\begin{equation}
\frac{E_{MF}}{N} = \sum_{i>j} \left(
J_{ij}  \left| Q_{ij} \right|^2
- J_{ij} Q_{ij} \varepsilon_{\sigma\sigma^{\prime}}
 x_i^{\sigma} x_{j}^{\sigma^\prime} + \mbox{H. c.} \right)
-\sum_{i} \lambda_{i} \left( 1+ \frac{n_b}{N}
-|x_i^{\sigma}|^2 \right)
+ \sum_{\mu} \omega_{\mu} ( Q , \lambda )
\label{emf}
\end{equation}
In the last term we have emphasized that the
$\omega_{\mu}$ depend upon  $Q, \lambda$.
Finding the ground state of $H_{AF}$ in the large $N$ limit
is
now
reduced to the problem of minimizing $E_{MF}$ with respect
to
the independent variables $Q_{ij}$, $x_{i}^{\sigma}$ with
the
$\lambda_i$ chosen such that the constraints
\begin{equation}
\frac{\partial E_{MF}}{\partial \lambda_i} = 0
\label{constraint}
\end{equation}
are always satisfied. It is instructive to examine the
equations obtained by demanding stationarity of $E_{MF}$
w.r.t
$x_{i}^{\sigma}$:
\begin{equation}
\sum_{j} \varepsilon_{\sigma\sigma^\prime} J_{ij} Q_{ij}
x_{j}^{\sigma^\prime}
+ \lambda_i x_{i\sigma}^{\ast} = 0
\label{stx}
\end{equation}
This equation has two possible solutions: {\em (i)}
$x_{i}^{\sigma} = 0$:
this gives the SRO phases
{\em (ii)} $x_{i}^{\sigma} \neq 0$: comparing Eqn
(\ref{stx}) with
Eqn (\ref{eig}) we see that this condition is equivalent to
demanding
that Eqn (\ref{eig}) possess at least one zero eigenvalue.
For finite $n_b$, this will only occur in the
limit $N_{s} \rightarrow \infty$, and implies the existence
of gapless
excitations. These are the Goldstone modes associated with
the $Sp (N)$ symmetry breaking.

We now examine the large $n_b / N$ limit of
$H_{MF}$;
it is easy to show that $Q_{ij} \sim \lambda_i \sim n_b /N$
while $x_{i}^{\sigma} \sim \sqrt{n_b /N}$. From Eqn
(\ref{emf})
we observe that the sum over $\omega_{\mu}$ can be neglected
and
for large $n_b / N$, $E_{MF}$ reduces to
\begin{equation}
\frac{E_{MF}^{c}}{N} = \sum_{i>j} \left(
J_{ij} \left| Q_{ij} \right|^2
- J_{ij} Q_{ij} \varepsilon_{\sigma\sigma^{\prime}}
 x_i^{\sigma} x_{j}^{\sigma^\prime} + \mbox{H. c.} \right)
-\sum_{i} \lambda_{i} \left( \frac{n_b}{N}
-|x_i^{\sigma}|^2 \right)
\end{equation}
The minimization of $E_{MF}^{c}$ w.r.t. $Q_{ij}$ can now be
easily
carried out and the problem reduces to that of
finding the classical ground state of $H_{AF}$. Thus the
large
$n_b / N$
limit of the present large $N$ equations is
equivalent to
the classical limit which involves
$n_b \rightarrow \infty$ at fixed $N$. The fact that these
two limits
commute gives us further confidence on the usefulness of the
present
large $N$ procedure.

Finally we indicate how this formalism can be used to
evaluate the
spin-stiffness of the LRO phases in the large $N$ limit. One
takes
values of $x_i^{\sigma}$ which minimize $E_{MF}$ and
performs
a slowly varying rotation:
\begin{equation}
x_{i}^{\sigma} \rightarrow \left(\exp\left(i \frac{\vec{n}
\cdot \vec{\tau}}{2} \vec{k} \cdot \vec{R}_i
\right)\right)_{\sigma
\sigma^\prime} x_{i}^{\sigma^\prime}
\end{equation}
where $\vec{n}$ is the direction in spin-space about which
the
rotation has been performed, $\vec{\tau}$ are the Pauli
matrices,
and $\vec{k}$ is a small wavevector. We insert this value
of $x_i^{\sigma}$ in $E_{MF}$, reminimize w.r.t. the
$Q_{ij}$
while maintaining the constraints and determine the change in
energy $\Delta E_{MF}$ for small $\vec{k}$. We expect in
general
\begin{equation}
\Delta E_{MF} = \frac{N_s}{2} \rho_{\alpha\beta} ( \vec{n}
) k_{\alpha} k_{\beta}
+ \cdots
\end{equation}
The co-efficient
$\rho$
is the spin-stiffness tensor of the LRO phase about the
direction $\vec{n}$.

\section*{\Large\bf 3. The ${\bf J_1}$-${\bf J_2}$-${\bf
J_3}$ Model}
We now apply the formalism developed in the previous section
to a frustrated antiferromagnet on the square lattice with
first ($J_1$), second ($J_2$) and third ($J_3$) neighbor
interactions. This model has been examined by
numerical~\cite{elbio,numj3}, series~\cite{rajiv} and
mean-field~\cite{rnb} methods in
the literature; their results will be compared with ours in
section 3.D. We begin by presenting the results of the
$N=\infty$ mean field theory (section 3.A). The effective
actions controlling the long-wavelength, long-time
fluctuations at finite $N$ will be considered in section
3.B. Berry phases and the spin-Peierls order they
induce
will be disussed in section 3.C.

\section*{\large\bf 3.A Mean Field Theory}

The energy $E_{MF}$ for this model was
minimized over all fields $Q_{ij}$ and $\lambda_i$ which were periodic
with a $\sqrt{2}\times \sqrt{2}$ unit cell. All of the
global minima found in fact turned out to have a periodicity
with one site per unit cell: we will therefore restrict our
analysis to this simpler limit. In addition, full
wavevector-dependent stability matrices involving
quadratic
fluctuations of $E_{MF}$ on the manifold specified by
Eqn (\ref{constraint}) were evaluated at representative
points,
and all eigenvalues
found to be positive. This gives us reasonable confidence
that the states described below are in fact the global
minima of $E_{MF}$.

With one site per unit cell, the link variational parameters
are the nearest neighbor fields $Q_{1,x}, Q_{1,y}$,
the diagonal 2nd neighbor fields $Q_{2,y+x}, Q_{2,y-x}$
and the 3rd neighbor fields $Q_{3,x}, Q_{3,y}$. The
bosonic eigenmodes can be calculated exactly and we find
the eigenenergies:
\begin{displaymath}
\omega_{\bk} = \left( \lambda^2 - 4|A_{\bk}|^2 \right)^{1/2}
\end{displaymath}
\begin{displaymath}
A_{\bk} = J_1 \left( Q_{1,x}\sin k_x + Q_{1,y}\sin k_y
\right)
+J_2 \left(
Q_{2,y+x}\sin (k_y + k_x ) +Q_{2,y-x}\sin (k_y - k_x )
\right)
\end{displaymath}
\begin{equation}
~~~~~~~~~~~~+J_3 \left(
 Q_{3,x}\sin (2k_x ) + Q_{3,y}\sin (2 k_y ) \right)
\label{omegak}
\end{equation}
The wavevector $\bk$ extends over the first Brillouin zone
of the square lattice.

Our results for $J_3 = 0$ are summarized in Fig~\ref{fig1}.
We begin by discussing the structure of the ground state
at $N=\infty$.
We find 3 distinct types of phases:

\subsubsection*{\bf 1. ${\bf (\pi , \pi )}$}
These states have $Q_{1,x}=Q_{1,y}\neq 0 $,
$Q_{2,y+x}=Q_{2,y-x}=Q_{3,x}=Q_{3,y}=0$
and are the analogs of the states found in $SU(N)$ systems.
Inserting these values into Eqn (\ref{omegak}) we see
that $\omega_{\bk}$ has its minima at $\pm (\pi/2 ,
\pi/2)$:
this implies that the spin-spin correlation function,
which involves the product of two bosonic pair correlation
functions~\cite{assa}, will have a peak at $(\pi , \pi)$.
We have a state with $x_{i}^{\sigma} \neq 0$ (LRO)
for $N/n_b < 2.5$ and the corresponding SRO
state for $N/n_b > 2.5$. The boundary between LRO and
SRO is independent of $J_2 / J_1$, but this is surely
an artifact of the large $N$ limit.
Finite $N$ fluctuations should be stronger as
$J_2 / J_1$ increases, causing the boundary to bend a little
downwards to the
right. While properties of these states can be calculated
numerically for all values of $J_2 / J_1$ and $N/ n_b$,
we tabulate a few analytic results for the LRO phase
in an expansion in powers of $N/n_b$.
The ground state energy is
\begin{equation}
E_{MF} = - N N_s \left( \frac{n_b }{N} \right)^2
\left[ 2J_1 + 4(1-I_1)J_1 \left(\frac{N}{n_b} \right )
+ \cdots \right]
\label{epp}
\end{equation}
where $I_1 = 0.84205$ is obtained from an integral over
the Brillouin zone. The spin-stiffness for spin rotations
about an axis perpendicular to the \neel axis is found to be
\begin{equation}
\rho_{\alpha\beta} = \delta_{\alpha\beta}
N \left( \frac{n_b }{N} \right)^2
\frac{1}{2}\left[
J_1 - 2J_2 + \left[
4J_2 (I_2 -1 ) - J_1 ( I_1 + I_2 -2) \right]
\left(\frac{N}{n_b} \right ) + \cdots \right]
\label{rpp}
\end{equation}
with $I_2 = 1.39320$.

\subsubsection*{\bf 2. ${\bf (\pi , 0)
}$ or ${\bf (0, \pi )}$}
The $(0,\pi)$ states have $Q_{1,x}=0$, $Q_{1,y} \neq 0$,
$Q_{2,y+x} = Q_{2,y-x} \neq 0$, and $Q_{3,x} = Q_{3,y } =0$
and occur with LRO and SRO. The degenerate
$(\pi , 0)$ state is obtained with the mapping $x
\leftrightarrow y$.
For $J_2 / J_1 > 1/2$, the classical limit,
has independent N\'{e}el order on each of the $A$ and $B$
sublattices; quantum
fluctuations, which are automatically included in the
present approach,
cause the N\'{e}el order parameters to align (``order from
disorder'')
leading to the LRO states.
To understand the nature of the transition between
LRO at $(\pi,\pi)$ to LRO at $(\pi,0)$ or $(0,\pi)$
we need the energy and stiffness of the latter states
for small $N/n_b$. We find
\begin{equation}
E_{MF} = - N N_s \left( \frac{n_b }{N} \right)^2
\left[ J_1 + 2J_2  + \left(
2J_1 + 4J_2 - 4J_1 I_3  \right)
 \left(\frac{N}{n_b} \right )
+ \cdots \right]
\label{epz}
\end{equation}
where $I_3$ is a Brillouin zone integral whose value depends
upon $J_2 / J_1$; we
will only need its value at $J_2 / J_1 = 0.5$: $I_3  =
0.88241$.
The stiffness of the
$(0 , \pi )$ state is soft at small $n_b / N$ for
wavevectors in the $x$-direction:
\begin{equation}
\rho_{xx} = N \left( \frac{n_b }{N} \right)^2
\frac{1}{2}\left[
2J_2 - J_1 + 2\left[
J_2 (2+I_5 -2I_4 ) + J_1 ( I_4 - 1) \right]
\left(\frac{N}{n_b} \right ) + \cdots \right]
\label{rpz}
\end{equation}
where $I_4 , I_5$ are Brillouin zone integrals dependent
upon $J_2 / J_1$. Again we will only need their values
at $J_2 / J_1 = 0.5$ which are $I_4 = 1.28576$ and $I_5 =
0.32767$.
Comparing Eqns (\ref{epp}) and (\ref{epz}) we see that the
energies of the $(\pi,\pi)$-LRO and $(0,\pi)$-LRO states
become equal at
\begin{equation}
J_{2} = J_1 \left[ \frac{1}{2} + 0.08072 \left(\frac{N}{n_b}
\right )
+ \cdots \right]
\end{equation}
At this value of $J_{2}/J_{1}$ the stiffness of the $(\pi ,
\pi)$
state is found to be
\begin{equation}
\rho_{\alpha\beta} (\pi , \pi ) = 0.1949
N \left( \frac{n_b }{N} \right)  \delta_{\alpha\beta} +
\cdots
\end{equation}
while the stiffness of the $(0, \pi)$ state is
\begin{equation}
\rho_{xx} (0 , \pi ) = 0.2446
N \left( \frac{n_b }{N} \right)  + \cdots
\end{equation}
Thus both stiffnesses are positive and indicate that the
transition between the states is first order.
Note that in both stiffnesses the leading $N(n_b / N)^2 $ term is absent;
these stiffnesses thus vanish classically and
have appeared directly as a result of
quantum fluctuations---another example of ``order from
disorder''.
These results contradict
arguments in Ref.~\cite{benoit} that another `spin liquid'
phase
exists at arbitrarily small $N/n_b$ near
$J_2/J_1 = 0.5$.
While we agree that quantum fluctuations are very important
in a `fan'
emanating from $J_2/J_1 = 0.5$, $N/n_b = 0$,
we have found that these fluctuations just reinforce the
classical order. At larger values of $N/n_b$ we
have a continuous (at $N=\infty$) transition to
the $(0,\pi)$-SRO or $(\pi , 0)$-SRO state. The transition
between this state and $(\pi , \pi )$-SRO remains
first-order.

\subsubsection*{\bf 3. ``Decoupled''}
For $J_2 / J_1 $  and $N/n_b$ both
large, we have a ``decoupled'' state
with $Q_{2,y+x} = Q_{2,y-x} \neq 0$ and
$Q_1=Q_3=0$. In this case
$Q_p$ is non-zero only for sites on the same
sublattice. The two sublattices have \neel type SRO
which will be coupled by finite $N$ fluctuations.
The $N=\infty$
state does not break any lattice symmetry.

We now turn to $J_3 \neq 0$ (Figs~\ref{fig2}, \ref{fig3}, \ref{fig4}),
where we find a new class of phases:

\subsubsection*{\bf 4. Incommensurate phases}
\begin{figure}
\centerline{\includegraphics[width=7.5in]{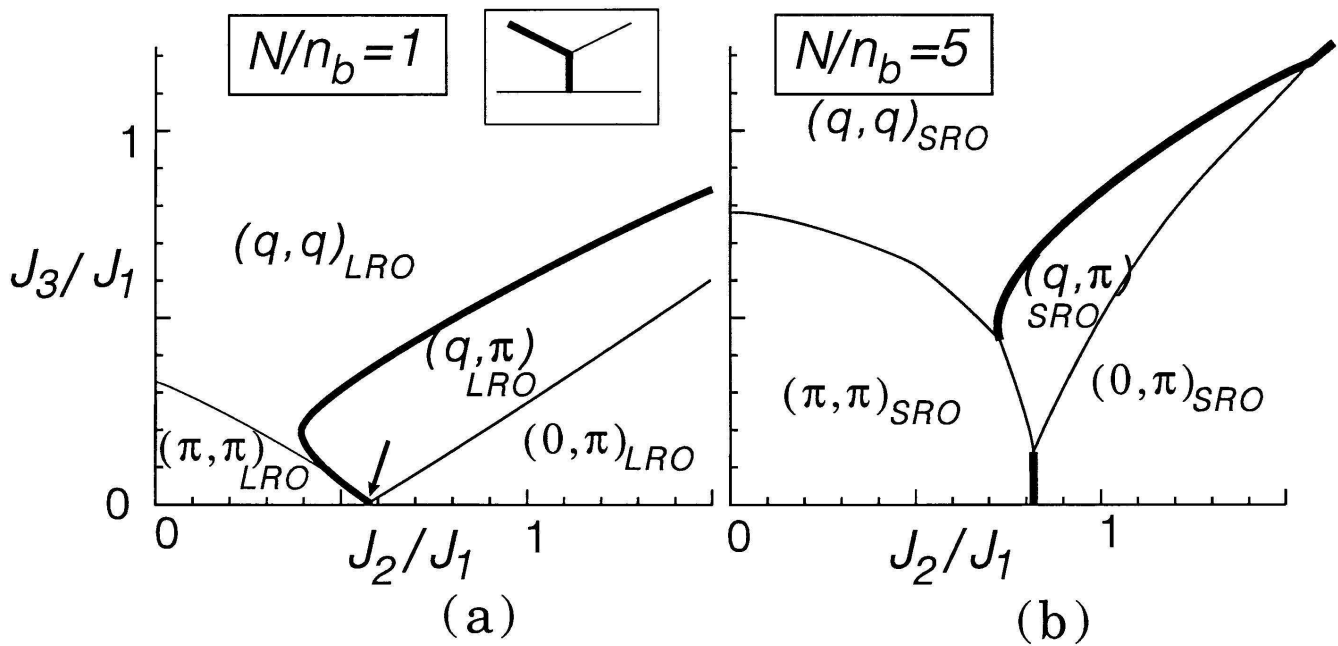}} \caption{ As
in Fig~\protect\ref{fig1} but as a function of $J_2 / J_1$ and
$J_3/ J_1$ for ({\it a\/}) $N/n_b = 1$ and ({\it b\/}) $N/n_b =5$.
The inset in ({\it a\/}) shows the region at the tip of the arrow
magnified by 20: a direct first-order transition from $(\pi ,
\pi)$-LRO to $(0 , \pi)$-LRO occurs up to $J_3/ J_1 = 0.005$.}
\label{fig2}
\end{figure}

\begin{figure}
\centerline{\includegraphics[width=7.5in]{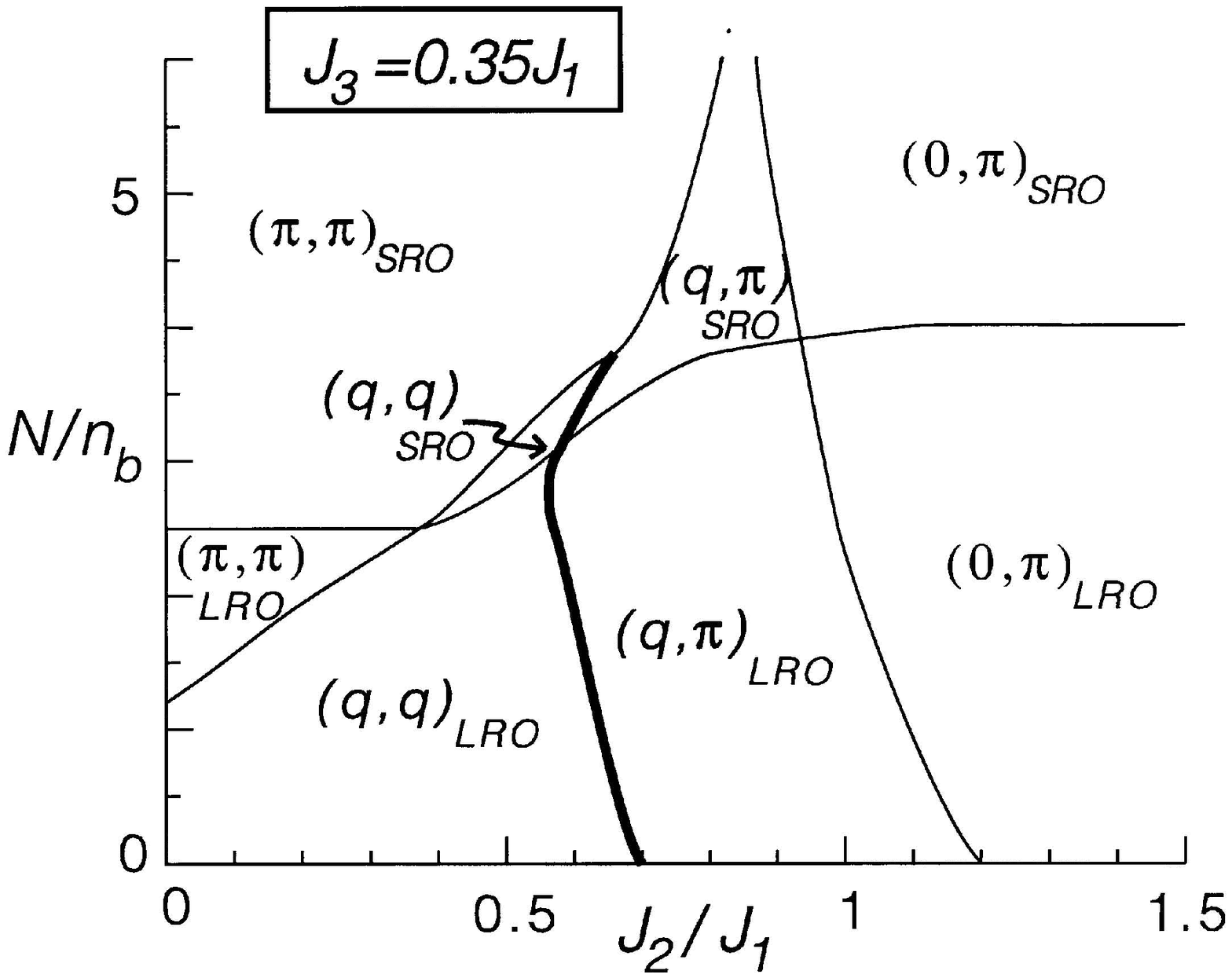}} \caption{ As
in Fig~\protect\ref{fig1} but for $J_3 / J_1 = 0.35$.}
\label{fig3}
\end{figure}

It is known that for finite $J_3$, the classical
magnet has phases with incommensurate helical (coplanar) order~\cite{rajiv}.
Our phase diagram for a small value of $N/n_b$
($N/n_b = 1$, Fig~\ref{fig2}a) is similar to the classical one.
All of the LRO phases undergo continuous transitions
to SRO with the same spatial distribution of the
$Q_p$ and the accompanying
broken rotational symmetry of the lattice.
These SRO phases are shown in Fig~\ref{fig2}b. The two new
classes of
phases which did not appear at $J_3 = 0$ are:
({\it i\/})
$\pm (q , \pi )$ phases with
$Q_{1,x}\neq Q_{1,y} \neq 0$,
$Q_{2,x+y} = Q_{2,y-x} \neq 0$, $Q_{3,x} \neq 0$ and
$Q_{3,y}=0$; the
degenerate $\pm(\pi,q)$ helix is obtained by the mapping $x
\leftrightarrow
y$.
({\it ii\/})
$\pm (q , q)$ phases with $Q_{1,x}=Q_{1,y} \neq 0$,
$Q_{2,x+y}\neq 0$, $Q_{2,y-x} = 0$, $Q_{3,x}=Q_{3,y}\neq 0$;
this
is degenerate with $(q, -q)$ phases\cite{shraisiggkane}.
The wavevector $q$ varies
smoothly from $0$ to $\pi$ and is
determined by doubling the wavevector at which
$\omega_{\bk}$ has a minimum; it is continuous across
second-order
phase boundaries (Fig~\ref{fig2}a).
In contrast to the classical limit~\cite{rajiv},
for $N/n_b$ finite it requires a finite $J_3$ to induce
helical order;
the first order transition from
$(\pi,\pi)$ to $(\pi,0)$ order persists for small $J_3$
as shown in the inset of Fig~\ref{fig2}a. This suggests that quantum
fluctuations induce an effective ferromagnetic $J_3$
in the model with bare $J_3 = 0$; a finite bare
antiferromagnetic $J_3$ is required to compensate for this.

The broken discrete symmetries in states
with SRO at $(0 , \pi )$ and $(q , \pi)$ are identical:
both are two-fold degenerate due to a breaking of the
$x \leftrightarrow y$ symmetry. The states
are only distinguished by a non-zero value of $Q_3$
in
the $(q, \pi )$ phase and the accompanying incommensurate
correlations in the spin-spin correlation functions.
However $Q_3$ is gauge-dependent
and so somewhat
unphysical as an order parameter.
In the absence of any further fluctuation-driven lattice
symmetry
breaking (see below), the transition between SRO at $(0, \pi
)$
and $(q , \pi)$ is an example of a {\em disorder
line}~\cite{disorder}; these are lines at which
incommensurate correlations first turn on.

Fig~\ref{fig3} shows the transition from LRO to SRO as a function of
$N/n_b$ for $J_3 = 0.35 J_1$.
Similar transitions also appear in Fig~\ref{fig4} which has
$J_3 = J_2 / 2$. There are several notable features of
these two phase diagrams:
\newline
({\it i\/})
An intermediate state in which $Sp(N)$ symmetry is only
partially
restored (spin-nematic~\cite{premi}) does not appear in the
large $N$
limit. We expect the transition from $(q, q)$-LRO
to $(q,q)$-SRO
to be described by a $NL\sigma$ model on the manifold
$M_{\rm noncoll}$.
A $d=1+\epsilon$ expansion of a similar $NL\sigma$
model~\cite{joli} for $N=1$ has been recently carried out
and
does not show any tendency towards such ordering.
\newline
({\it ii\/})The $Q_p$ variables can all be
chosen real in all the
phases,
indicating the absence of ``chiral''~\cite{wwz} order.
\newline
({\it iii\/}) The
commensurate
states squeeze out the incommensurate phases as $N/n_b$
increases in both phase diagrams.
We expect that this suppression of incommensurate order by
quantum
fluctuations is
a general feature of frustrated
AFMs.

\section*{\large\bf 3.B Fluctuations - long wavelength
effective actions}
This section will examine the form of the effective action
controlling the long-wavelength fluctuations of the
$b^{\alpha}$
quanta and the link fields $Q_p$. We will focus mainly
on the SRO phases in the region close to the transition to
the
LRO phases.

We begin by examining the $(\pi , \pi)$-SRO phase; this
phase
has already been examined in considerable detail in
Ref~\cite{self2}
for the $SU(N)$ AFM; here we adapt the results to the
$Sp(N)$ AFM. As noted above, this phase has the mean field
values
$Q_{1,x} = Q_{1,y} = \bqo$, and $\omega_{\bk}$ has minima at
$\pm \vec{G}$, with $\vec{G} = (\pi/2 , \pi/2)$. Close to
the LRO
phases, fluctuations of the $b^{\alpha}$ will be
strong
at $\pm \vec{G}$. We therefore parametrize
\begin{equation}
b_{i}^{\alpha} = \varphi_{+}^{\alpha} ( \vec{r}_i ) e^{i
\vec{G} \cdot
\vec{r_i}} + \varphi_{-}^{\alpha} ( \vec{r}_i ) e^{-i
\vec{G} \cdot
\vec{r_i}}
\label{ansb}
\end{equation}
where $\varphi_{\pm}$ are assumed to have slow spatial
variations.
The most important component of the fluctuations of the
$Q_{ij}$
are those of their phases. In particular, as noted in
Ref~\cite{self2},
the mean-field theory has an unbroken gauge symmetry; the
transformation
\begin{equation}
b_{i}^{\alpha}  \rightarrow  b_{i}^{\alpha} e^{i \gamma_i
\theta}
\label{gtrans}
\end{equation}
where $\gamma_i = 1 (-1)$ for
$r_{ix} + r_{iy}$ even (odd),
leaves all $\left\langle Q_{ij} \right\rangle$ invariant. A
staggered component of the phases of the $Q_{ij}$ acts as
the vector
potential associated with this gauge symmetry. We therefore
write
\begin{equation}
Q_{i,i+ \hat{x}} = \bqo \exp \left( i \gamma_i A_{x}
(\vec{r}_i + \hx /2)
\right)
\label{ansq1}
\end{equation}
where again $A_{x}$ is slowly varying; a similar equation
holds
for $A_{y}$. The time-component of the gauge-field is the
fluctuation
in the Lagrange multiplier:
\begin{equation}
\lambda_i = \bar{\lambda} + i\gamma_i A_{\tau} ( \vec{r}_i )
\label{anslam}
\end{equation}
We now insert the ansatzes (\ref{ansb}), (\ref{ansq1}) and
(\ref{anslam}) into
the Lagrangian ${\cal L}$
associated with $H_{AF}$ and perform a gradient expansion
on all the terms. The final result is most compactly
expressed in
terms of the fields
\begin{eqnarray}
\psi_1^{\alpha} &=& ( \varphi_{+}^{\alpha} + \varphi_{-
}^{\alpha})/
\sqrt{2}\nonumber \\
\psi_{2\alpha} & =& -i \cj_{\alpha\beta} (
\varphi_{+}^{\beta} - \varphi_{-}^{\beta} ) /\sqrt{2}
\end{eqnarray}
and has the form
\begin{displaymath}
{\cal L} = \int \frac{d^2 r}{a^2} \left [
\psi_{1\alpha}^{\ast} \left( \frac{d}{d\tau} + i A_{\tau}
\right)
\psi_1^{\alpha} +
\psi_{2}^{\alpha\ast} \left( \frac{d}{d\tau} - i A_{\tau}
\right)
\psi_{2\alpha} + \bar{\lambda} \left( |\psi_1^{\alpha} |^2
+ |\psi_{2\alpha} |^2 \right) \right.
\end{displaymath}
\begin{equation}
-4 J_1 \bqo \left ( \psi_1^{\alpha}\psi_{2\alpha} +
\psi_{1\alpha}^{\ast}\psi_2^{\alpha\ast}
\right )
+  J_1 \bqo a^2 \left [
\left ( \vec{\nabla}  + i  \vec{A} \right ) \psi_1^{\alpha}
\left ( \vec{\nabla}  - i \vec{A} \right ) \psi_{2\alpha}
+ \left ( \vec{\nabla} - i \vec{A} \right )
\psi_{1\alpha}^{\ast}
\left ( \vec{\nabla} + i \vec{A} \right )
\psi_2^{\alpha\ast} \right ]  \Biggr]
\label{charge}
\end{equation}
We now introduce the fields
\begin{eqnarray*}
z^{\alpha} & = & (\psi_1^{\alpha} +
\psi_2^{\alpha\ast})/\sqrt{2} \\
\pi^{\alpha} & = & (\psi_1^{\alpha} -
\psi_2^{\alpha\ast})/\sqrt{2} .
\end{eqnarray*}
From Eqn (\ref{charge}), it is clear that the
the $\pi$ fields turn out to have mass $\blam + 4 J_1 \bqo$,
while the $z$ fields
have a mass $\blam - 4 J_1 \bqo$ which vanishes at the
transition to the LRO
phase. The $\pi$ fields can therefore
be safely integrated out,
and ${\cal L}$ yields
the following effective action, valid at distances much
larger than the lattice
spacing:
\begin{equation}
S_{eff} =
\int \frac{d^2 r}{\sqrt{8}a} \int_{0}^{c\beta}
d\ttau \left\{
|(\partial_{\mu} - iA_{\mu})z^{\alpha}|^2
+ \frac{\Delta^2}{c^2}
|z^{\alpha} |^2\right\},
\label{sefp}
\end{equation}
Here $c = \sqrt{8}J_1 \bqo a$ is the spin-wave velocity,
$\Delta = (\lambda^2 - 16J_1^2 \bqo^2 )^{1/2}$ is the gap
towards spinon excitations,
and $A_{\ttau} = A_{\tau}/c$; this action is
of the form $(\ref{sefpp})$ quoted in the introduction.
Thus, in its final form, the long-wavelength theory consists
of a massive $z^{\alpha}$ scalar (spinon)
coupled to a compact $U(1)$ gauge field.

We now examine the changes in the above actions as one moves
from the $(\pi , \pi)$-SRO phase into the $(q,q)$-SRO phase
(Fig~\ref{fig2}b) (a very similar analysis can be performed at the
boundary between the $(\pi , \pi)$-SRO and the $(\pi , q)$-
SRO phases). This transition is characterized by a
continuous
turning on of non-zero values of $Q_{i,i+\hy+\hx}$,
$Q_{i,i+2\hx}$ and $Q_{i,i+2\hy}$. It is easy to see from
Eqn (\ref{gtrans}) that these fields transform as scalars of
charge $2\gamma_i$ under gauge transformation associated
with $A_{\mu}$. Performing a gradient expansion upon the
bosonic fields coupled to these scalars we find that the
Lagrangian ${\cal L}$ of the $(\pi , \pi)$-SRO phase gets
modified to
\begin{equation}
{\cal L} \rightarrow {\cal L} + \int \frac{d^2 r}{a} \left(
\vec{\Phi}_A \cdot \left(
\cj_{\alpha\beta}\psi_1^{\alpha}\vec{\nabla}\psi_1^{\beta}
\right) +
\vec{\Phi}_B \cdot \left(
\cj^{\alpha\beta}\psi_{2\alpha}\vec{\nabla}\psi_{2\beta}
\right) + \mbox{c.c.} \right)
\end{equation}
where $\vec{\Phi}_{A,B} \equiv (J_3 Q_{3,x} + J_2 Q_{2,y+x},
J_3 Q_{3,y} + J_2 Q_{2,y+x} )$ with the sites on the ends of
the link
variables on sublattices $A,B$. Finally, as before, we
transform to the $z,\pi$ variables, integrate out the $\pi$
fluctuations and obtain
\begin{equation}
S_{eff} =
\int \frac{d^2 r}{\sqrt{8}a} \int_{0}^{c\beta}
d\ttau \left\{
|(\partial_{\mu} - iA_{\mu})z^{\alpha}|^2
+ \frac{\Delta^2}{c^2}
|z^{\alpha} |^2 + \vec{\Phi} \cdot \left(
\cj_{\alpha\beta}z^{\alpha}\vec{\nabla} z^{\beta}\right) +
\mbox{c.c.}
+ V(\Phi) \right\} + \dots,
\label{hgs}
\end{equation}
Here $\vec{\Phi} = (\vec{\Phi}_A + \vec{\Phi}_B^{\ast} )
/(2J_1 \bqo a)$ is a
scalar of
charge $-2$; terms higher order in $\vec{\Phi}$ have been dropped.
We have also added a phenomenological potential
$V(\Phi )$ which is generated by short wavelength
fluctuations of the $b^{\alpha}$ quanta.
This action is of the general form (\ref{hgss}),
and as noted earlier, is the simplest one consistent
with the requirements of $U(1)$-gauge
and global $Sp(N)$ invariance.
In a phase with $\left\langle\vec{\Phi}
\right\rangle \neq 0$ and real, we
see from Eqn (\ref{hgs}) that the minimum of the dispersion
of the $z^{\alpha}$ quanta is at a wavevector $\bk_0 =
(\langle\Phi_x\rangle
, \langle\Phi_y\rangle )/2$:
this will lead to incommensurate
order. The action (\ref{hgs}) thus demonstrates
clearly the
intimate connection between the
Higgs phenomenon and the
appearance of incommensurate correlations.

We now turn to the $(0 , \pi)$-SRO phase; we will find that
its properties are quite similar to the $(\pi , \pi)$-SRO
phase.
We recall that this phase has the mean-field values
$Q_{2,y+x} = Q_{2,y-x} = \bar{Q}_2$, $Q_{1,y} = \bqo$ and
$Q_{1,x} = 0$. The excitation energies $\omega_{\bk}$
now have minima at $\pm \vec{G}$ with $\vec{G} = (0, \pi
/2)$. We therefore parametrize the $b^{\alpha}$ fields as in
Eqn (\ref{ansb}), but using the new value of $\vec{G}$.
This phase also has an unbroken $U(1)$ gauge symmetry;
the main difference is in the staggering pattern of the
charges of the $b^{\alpha}$ quanta. With the choice of the
constants $\gamma_i = 1 (-1)$ for $r_{iy}$ even (odd) in Eqn
(\ref{gtrans}) it is easily verified that all mean-field
expectations $\left\langle Q_{ij} \right\rangle$ remain
invariant. The phases of the link fields are parametrized as
in Eqn (\ref{ansq1}) in terms of a vector potential $A$ but
with the new $\gamma_i$; the diagonal link fields have the
parametrization
\begin{equation}
Q_{i,i+\hy+\hx} = \bar{Q}_2 \exp \left[ i \gamma_i \left(
A_{y} ( \vec{r}_i + \hy /2 + \hx /2 ) +
A_{x} ( \vec{r}_i + \hy /2 + \hx /2 ) \right) \right]
\end{equation}
and similarly for $Q_{i,i+\hy-\hx}$.
The remaining analysis is essentially identical to that for
the $(\pi , \pi)$-SRO phase and we obtain the same final
effective action $S_{eff}$ (Eqn (\ref{sefp})) with an
additional spatial anisotropy. The
connection between the $z^{\alpha}$ quanta and the lattice
bosons and between $A_{\mu}$ and the phases of the link
fields are now of course different.

The transition from $(0 , \pi)$-SRO to $(q , \pi)$-SRO
follows the treatment above of the transition from $(\pi ,
\pi)$-SRO to $(q , q)$-SRO; the final action for the
incommensurate phase has the same form as Eqn (\ref{hgs}).

\section*{\large\bf 3.C Berry Phases}
For the unfrustrated $SU(N)$ magnets, Berry phases
associated
with instanton tunneling events had particularly strong
consequences in the SRO phase~\cite{hedge,self2}.
Very similar effects occur
in the commensurate SRO phases of the $J_1$-$J_2$-$J_3$
model.
We consider the various phases in turn:
\subsubsection*{\bf 1. ${\bf (\pi , \pi )}$}
This phase is essentially identical to the $SU(N)$
case~\cite{self2}.
The instantons have Berry phases $n_b \zeta_s \pi /2$
where $\zeta_s = 0,1,2,3$ on dual lattice points with
(even,even), (even,odd), (odd,odd), (odd, ,even) co-
ordinates.
Condensation of the instantons
leads to spin-Peierls order
of
column (shown in Fig~\ref{fig1}) or line type (not
shown) for $n_b = 1,3 \mf$ and $2$ respectively,
or a featureless VBS state~\cite{aklt}
for $n_b = 0 \mf$, throughout the $(\pi,\pi)$ SRO phase.
The $b^{\alpha}$ quanta are confined by the compact $U(1)$
gauge force, with a confinement length scale determined by
the instanton density~\cite{polyakov}.
\subsubsection*{\bf 2. ``Decoupled''}
We can apply
the analysis of subsection 3.C.1 to each sublattice
separately, giving {\it e.g.\/} for $n_b = 1 \mf$
the type of spin-Peierls correlations shown in Fig~\ref{fig1}.
There is a total of $4\times 4 = 16$ states for this
case but coupling between the sublattices will
reduce this to 8 states, all of one of the two types shown.
The state with the `dimers' parallel to one another has
more possibilities for resonance using the $J_1$ bonds and
is likely to be the ground state.
For $n_b =
2 \mf$, there will be $2\times 2/2 = 2$ states, and for
$n_b = 0 \mf$, just one.

\subsubsection*{\bf 3. ${\bf (0, \pi )}$}
As in the other collinear phases, this state
possesses instantons which are the remnants of hedgehogs in
the
LRO phase. A natural location for these instantons is at the
center of the horizontal links which have
$Q_{1,x}=0$; the spins at the ends of these links are
ferromagnetically aligned and are most susceptible to large
deformations. The Berry phase for a configuration of
well separated instanton
charges $m_s$ at co-ordinates $\bR_s$ can be calculated as
before~\cite{self2,hedge} and we find
\begin{equation}
S_B = i n_b \sum_s \sum_i \left[ m_s \gamma_i \theta_s (\br_i )
\right]
\label{bersum}
\end{equation}
where $\theta_s (\br_i )$ is an angle which winds by
$2\pi$ as $\br_i$ moves around $\bR_s$,
the sum over $i$ extends over all the sites of the
lattice
and $\gamma_i$ is the staggering associated with the $(0 ,
\pi)$
phase {\em i.e.} $\gamma_i = 1 (-1)$ for $r_{iy}$ even
(odd).
The sum over $i$ can be evaluated in a manner similar to
Ref~\cite{hedge}. Consider first an isolated instanton of unit charge at
$\bR_s$; rewrite the summation in (\ref{bersum}) as one over
vertical links of the square lattice:
\begin{equation}
S_B = i n_b \sum_{i} \frac{\gamma_i}{2} \left[\theta_s (\br_i)
- \theta_s (\br_{i + \hy})\right]
\end{equation}
The symmetry of the mean-field $(0,\pi )$ state with one
instanton now requires that
the contribution of the links be
odd under
reflection across the horizontal line running through $\bR_s$.
The majority of the links will cancel against their reflection
partners, the only
exceptions being the links which intersect the cut in the angle
field $\theta_s (\br_i )$; these will yield an additional
contribution of $i\pi n_b$.
We expect that these symmetry considerations will also be valid for
a configuration of well separated instantons. Combining the contributions
of all the instantons we obtain finally
\begin{equation}
S_B = i \pi n_b \sum_s m_s \left[R_{sx}\right]
\end{equation}
where $\left[R_{sx}\right]$ is the integer part of $R_x$.
In the SRO phase the instantons interact with a Coulombic
$1/R$ potential; the instanton plasma can therefore be mapped
onto a dual
sine-Gordon model~\cite{self2} in which the instanton Berry phases appear
as frustrating phase-shifts in the arguments of the cosine
term~\cite{self2}. This model can be analyzed in a manner
which closely parallels Ref~\cite{self2}.
The Berry phases lead to condensation of the
instanton charges and
to spin-Peierls order of the type shown in
Fig~\ref{fig1} for $n_b$ odd, and a VBS state
for $n_b$ even.  Combined with the
choice $(0,\pi)$ or $(\pi,0)$ this gives
degeneracies $2,4,2,4$ for $n_b=0,1,2,3 \mf$.

\subsubsection*{\bf 4. Incommensurate Phases}
As the action (\ref{hgs}) makes clear, these phases
require an additional charge $-2$ field $\vec{\Phi}$
for their description. In the SRO phases, the $z^{\alpha}$
quanta can be integrated out and we are
left with a lattice
gauge theory of a compact $U(1)$ gauge field and a charge
$-2$
scalar.
The scalar fields can form
vortices with flux quantum $\pi$: instantons can change this
flux by $2\pi$, so only a $Z_2$ quantum number is
topologically
stable~\cite{bulbul}, in agreement with the analysis of the
LRO phase.
If the vortices proliferate
it will be necessary to consider Berry phases associated with them
and the instantons to understand the SRO phase. The following
phases
can occur in such a lattice gauge
theory~\cite{fradkin_shenker}:
\newline
({\it i\/})
A Higgs phase
in which the vortices and instantons are suppressed; Berry
phase
effects are therefore unimportant and this phase will be
insensitive to the precise value of $n_b$.
The $b^{\alpha}$ quanta carry
charge 1 and are unconfined~\cite{fradkin_shenker}: these
are spinons transforming under the fundamental of $Sp(N)$.
No such excitations were found in the commensurate phases.
Gauge excitations have a gap controlled
by the magnitude of the Higgs field $\left\langle \vec{\Phi}
\right\rangle$.
The incommensuration is
also controlled by $\left\langle \vec{\Phi} \right\rangle$, as has
been
noted earlier.
There is no breaking of lattice symmetries beyond those
found
at $N=\infty$. However with periodic boundary conditions
the ground state has an additional factor of 4 degeneracy
for all
$n_b$: the additional states are obtained by changing the
sign
of all $Q_{p}$ fields cut by a loop wrapped around the
system~\cite{bulbul}. This phase is expected to survive
in our phase diagram at finite $N$.
\newline
({\it ii\/}) A confinement phase with proliferation of
vortices
and instantons, confinement of spinons, and a gap towards
gauge excitations controlled by the instanton
density~\cite{polyakov}. A
plausible scenario is that this phase in fact coincides with
the $(\pi , \pi )$-SRO phase and possesses spin-Peierls
order driven by the Berry phases of instantons.
\newline
({\em iii\/}) Additional intermediate phases driven by the
Berry phases of the instantons and vortices---this
possibility
is under investigation.

\section*{\large\bf 3.D Comparison with numerical and series
results}

Many
numerical~\cite{elbio} and series
analyses~\cite{rajiv}
have appeared on the model with $J_3 = 0$ for the spin-1/2
$SU(2)$ model {\em i.e.} $N=1$, $n_b =1$ in the notation of
this paper. They find $(\pi , \pi )$-LRO at small $J_2 /
J_1$, $(\pi , 0)$-LRO at large $J_2 / J_1$ and an
intermediate SRO phase around $J_2 / J_1 = 1/2$. This is in
agreement (see Fig~\ref{fig1})
with our prediction in section 3.A.1 that the
phase boundary between $(\pi , \pi )$-LRO and $(\pi , \pi)$-
SRO bends downwards at finite $N$ with increasing $J_2 /
J_1$. Analyses of this intermediate
phase at $J_2 / J_1$~\cite{elbio,rajiv} shows clear evidence
of columnar spin-Peierls ordering~\cite{self2}, also in agreement
with the results of Ref~\cite{self2} and section 3.C.1.
An additional intermediate phase with $(0, \pi)$-SRO
has not been ruled out.

\begin{figure}
\centerline{\includegraphics[width=7.5in]{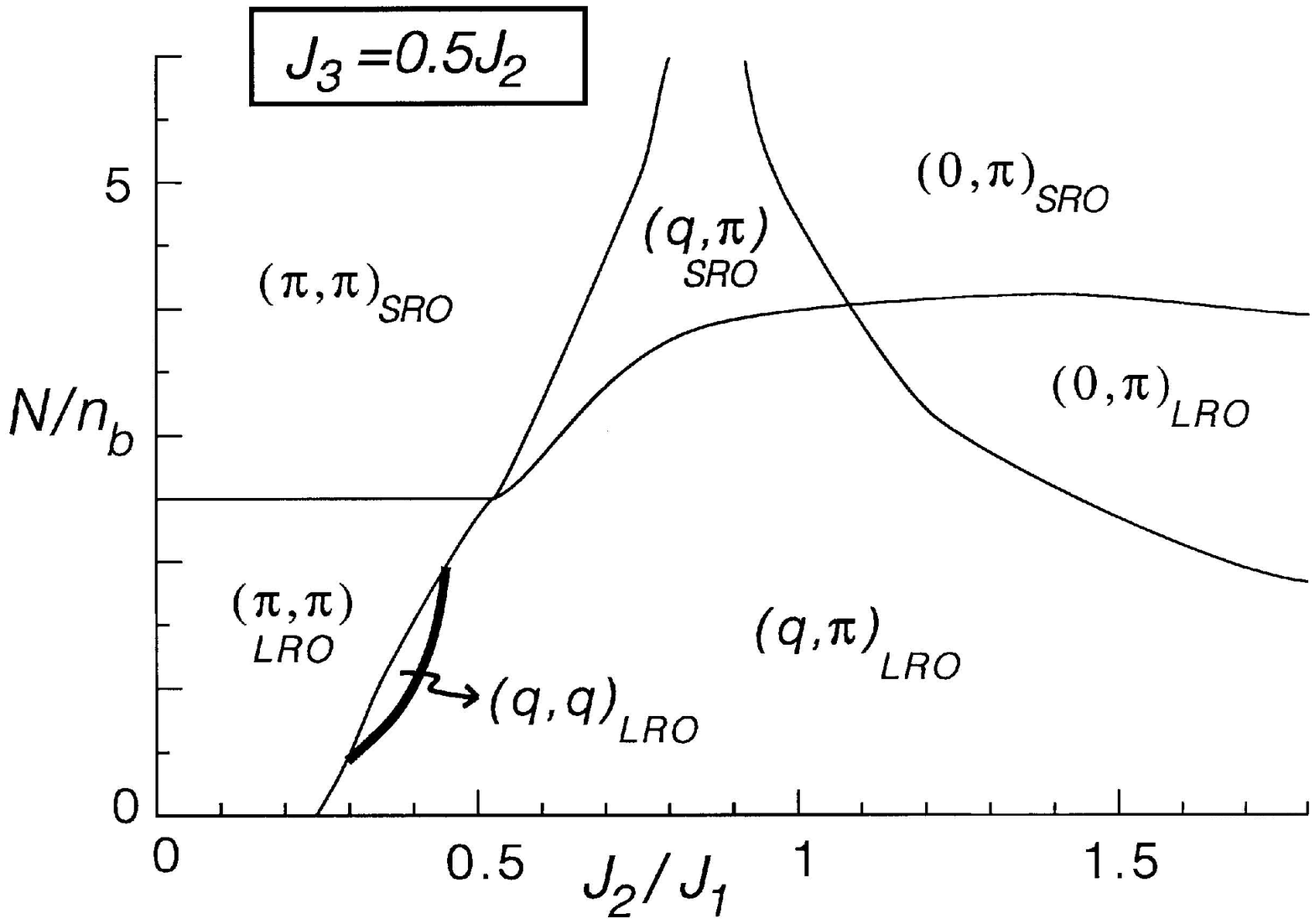}} \caption{ As
in Fig~\protect\ref{fig1} but for $J_3 / J_2 = 0.5$.} \label{fig4}
\end{figure}

It would clearly be interesting to find the new incommensurate
SRO phases with unconfined spinons of this paper in
numerical work on models with $N=1$, $n_b=1$. Our results
suggest that such phases will only occur in models with a
finite $J_3$. A numerical analysis of the $J_1$-$J_2$-$J_3$
model along the line $J_3 = 0.5 J_2$ has recently been
performed~\cite{numj3}. For comparison, we display the large
$N$ results for these models in Fig~\ref{fig4}.
As in Fig~\ref{fig4}, these investigators~\cite{numj3} find
correlations at $(\pi , \pi)$ at small $J_2 / J_1$ and at
$(0 , \pi)$ at large $J_2 / J_1$.
Their results at intermediate $J_2 / J_1$ are inconclusive
but suggestive of a $(\pi, q)$ phase. A systematic analysis
to resolve this issue and to distinguish SRO from LRO would
be useful.

\section*{\Large\bf 4. The ${\bf t}$-${\bf J}$ Model}

This section will extend our results to a limited class of
doped
AFMs. These are described by the $t$-$J$ models in which
electrons hop with matrix element $t$, have a
nearest-neighbor
antiferromagnetic exchange $J$, and have restrictions on
their
occupation number at each site due to the strong Coulomb
interactions.
Ideally one would like to dope AFMs which have spins in the
representations examined above, {\em i.e.} transforming
under
the symmetric product of $n_b$ fundamentals of $Sp(N)$, and
take
the large $N$ limit with $n_b / N$ constant. Parameters can
then
be chosen
to obtain the experimentally observed~\cite{sudip} \neel LRO
state in the undoped limit. These models require the
introduction of
bosons $b^{\alpha}$ carrying $Sp(N)$ spin, and spinless
fermions
to represent the holes~\cite{shraisiggkane}.
The Pauli exclusion principle now
restricts
the density of fermions to be order 1 and it appears that a
simple
large-$N$ limit does not exist.

We examine here the doping of AFMs which have spins
transforming under the antisymmetric product of $m$
fundamentals
of $Sp(N)$ (this representation is {\em not} irreducible);
the large $N$ limit will be taken with $m/N$
constant.
Such spins are most conveniently described by introducing
fermions $f^{\alpha}$ transforming under the fundamental of
$Sp(N)$. The holes are now represented by spinless bosons
$b$. The
physical `electron' is therefore
\begin{equation}
c_i^{\alpha} = f_i^{\alpha} b_i^{\dagger}
\end{equation}
The local constraint of the $t$-$J$ model is
\begin{equation}
f_{i\alpha}^{\dagger} f_i^{\alpha} + b_{i}^{\dagger} b_{i} =
m
\label{consf}
\end{equation}
for every site $i$. We will focus exclusively on the case
in which the states are half filled at zero doping: $m=N$.
This approach is an adaptation of the large $N$ limits of
Ref~\cite{jbm,gabi} for $SU(N)$ $t$-$J$ models.
Unlike the
$SU(N)$ case, the states on each site do not form an
irreducible
representation of $Sp(N)$ at zero doping. Two fermions can
combine to
form on-site singlets $\cj^{\alpha\beta} f_{i\alpha}^{\dagger}
f_{i\beta}^{\dagger}$. There is, however, no energy gained
out of
forming such states; they in fact reduce the ability of the
site
to form energetically favorable singlet bonds with other
sites.

For undoped AFMs with
exchange interactions of the type (\ref{hex}),
the large $N$ limit can be taken following the $SU(N)$ case.
For weak frustration~\cite{rok} the ground states are fully
dimerized into pairs of sites forming singlet bonds. (Note
that
the bosonic large $N$ theories of Section 3 also give only
dimerized
ground states in the limit of large $N/n_b$ for $n_b$ odd:
see Figs~\ref{fig1}, \ref{fig3}, \ref{fig4}.)
\neel-type states which are important for
$N=1$
are not found. However at large dopings, the present
large-$N$
offers a natural way of describing metallic states with a
Fermi surface
satisfying Luttinger's theorem, the existence of which
appears to be suggested by some experiments on the
high-$T_c$ materials~\cite{lutt}.
Thus the following calculation, while probably not
experimentally relevant at small doping, might capture the correct
physics at moderate and large dopings.

\begin{figure}
\centerline{\includegraphics[width=7.5in]{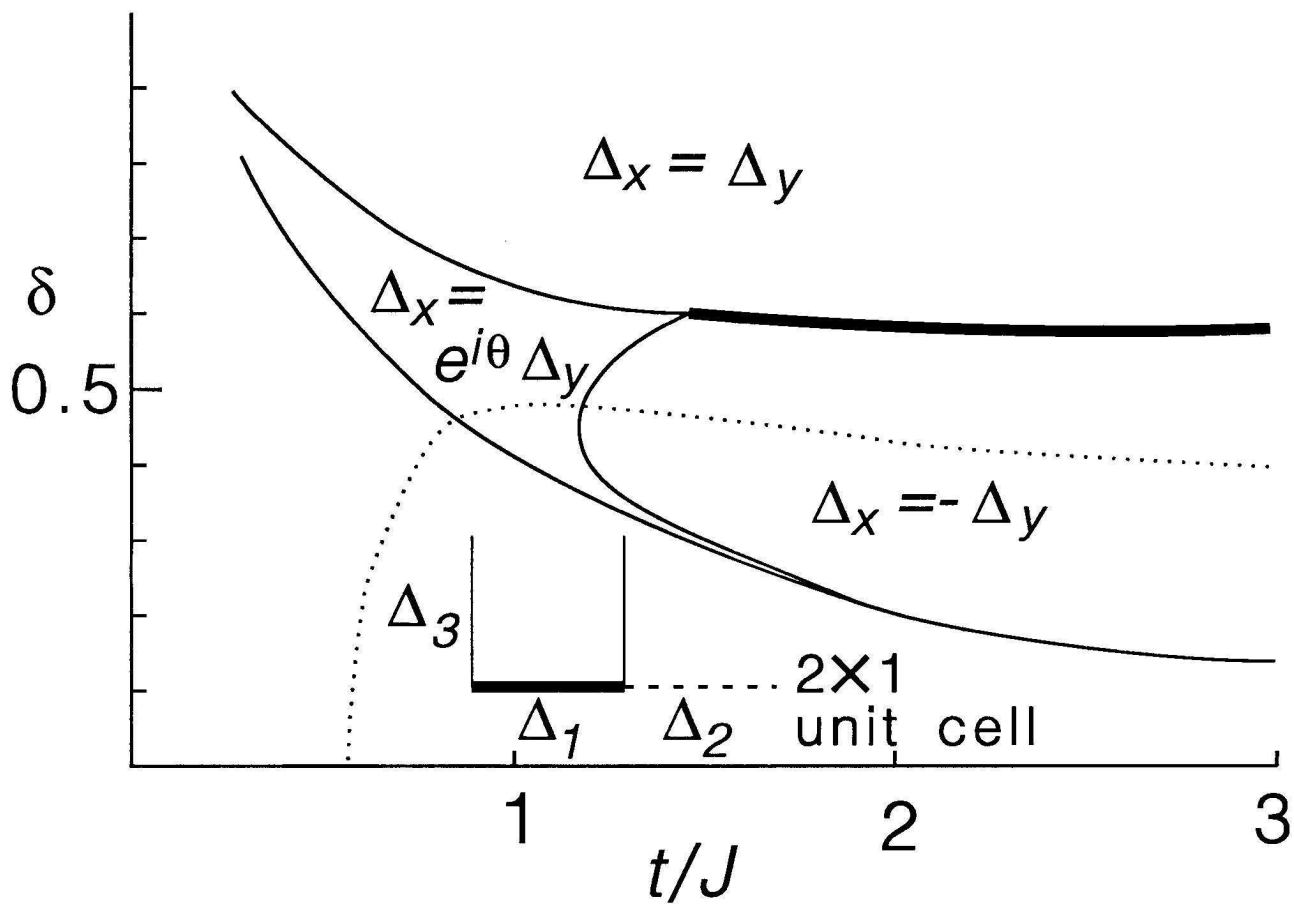}} \caption{
Ground state of the $t-J$ model $H_{tJ}$ at $N=\infty$ as a
function of $t/J$ and the hole concentration $\delta$. All phases
except on the lines $t=0$ and $\delta=0$ are superconducting. The
region below the dotted line is susceptible to separation into an
insulating antiferromagnet with $\delta = 0$ and a hole-rich phase
with $\delta$ on the dotted line. The boundary towards phase
separation approaches $\delta = 0 $ as $t/J \rightarrow \infty$}
\label{fig5}
\end{figure}

We will consider the following Hamiltonian
\begin{displaymath}
H_{tJ} = - \frac{t}{N} \sum_{<ij>}b_{i}
f_{i\alpha}^{\dagger}
f_j^{\alpha} b_{j}^{\dagger} - \frac{J}{N} \sum_{<ij>}
\left( \cj^{\alpha\beta} f_{i\alpha}^{\dagger}
f_{j\beta}^{\dagger} \right)
\left( \cj_{\gamma\delta} f_j^{\delta} f_i^{\gamma} \right)
\end{displaymath}
\begin{equation}
~~~~~~~~~~~~~~+ \sum_i \lambda_i \left(
f_{i\alpha}^{\dagger} f_i^{\alpha} + b_{i}^{\dagger} b_{i} -
N \right)
+ \mu \sum_i \left( b_i^{\dagger} b_i - N\delta \right)
\end{equation}
The last two terms enforce the local constraint
(\ref{consf})
and fix the average hole density at $N\delta$. Here we will
focus
exclusively on the $N=\infty$  limit at zero temperature.
This requires a complete condensation of the $b$ bosons
$\langle b_i \rangle = \sqrt{N} \bar{b}_i$ and introduction
of the
link-field
\begin{equation}
\Delta_{ij} = \frac{1}{N} \left\langle
\cj^{\alpha\beta} f_{i\alpha}^{\dagger}
f_{j\beta}^{\dagger}\right\rangle
\end{equation}
which performs the Hubbard-Stratanovich decomposition of the
exchange
interaction. We note that this decoupling is very similar to
that
performed in Refs~\cite{bza}; in the present calculation
however
it is uniquely enforced by the large $N$ limit.
As in Ref~\cite{bza}, at $T=0$,
our solutions have $\bar{b}_i \neq 0$ (for $\delta \neq 0$)
and
$\Delta_{ij} \neq 0$ which describe a superconducting
state.
We have examined mean-field equations numerically with
two sites per unit cell (with both the $2\times 1$
and $\sqrt{2} \times \sqrt{2}$ unit cells) and determined
the global
ground states over this manifold. The results are summarized
in
Fig~\ref{fig5}. A large part of the phase diagram in fact turns out
to have
a unit cell of just one site. In this case $\bar{b}_i =
\sqrt{\delta}$,
$\Delta_{i,i+\hx} \equiv \Delta_x$,
$\Delta_{i,i+\hy} \equiv \Delta_y$ and the quasi-particle
energy
spectrum $\epsilon_{\bk}$ is
\begin{equation}
\epsilon_{\bk} = \left[ \left( \lambda -2t\delta \left(
\cos k_x + \cos k_y \right) \right)^2 + 4 J^2 \left|
\Delta_x \cos k_x + \Delta_y \cos k_y \right|^2
\right]^{1/2}
\end{equation}

We find 4 different types of superconducting states which are described
below.
With the exception of the first phase which has coexisting spin-Peierls
order, all phases are combination of extended $s$ ($s^\ast$) and $d$ wave
pairing with no on-site pairing.
\newline
({\it i\/}) Coexistence of spin-Peierls order and
superconductivity:
This phase occurs for small $t$ or small $\delta$ and has
a $2\times 1 $ unit cell with the spatial distribution of
the
$\Delta_{ij}$ shown in Fig~\ref{fig5}. All three $\Delta_1 , \Delta_2
,
\Delta_ 3$ cannot be made real in any gauge so this phase
also breaks time-reversal invariance. At $t=0$ and
$\delta$ arbitrary, or $\delta=0$ and $t$ arbitrary, we have
$\Delta_1 \neq 0$, $\Delta_2 = \Delta_3 = 0$: the system is
now a
fully dimerized insulator.
The quasi-particle spectrum is found to have a gap over
the entire Brillouin zone. This is the only phase which has
a
unit cell larger than one site.
\newline
({\it ii\/}) $s^{\ast} + id$-wave superconductor:
Occurs around $t/J \approx 1$ and $\delta \approx 0.5$ and
has
$\Delta_x = e^{i\theta} \Delta_y$ where the phase
$\theta$ varies smoothly with $t/J$ and $\delta$.
Such a state has been considered previously by
Kotliar~\cite{sid}.
The quasi-particle spectrum is fully gapped and
time-reversal
symmetry is broken.
\newline
({\it iii\/}) $d$-wave superconductor: This phase occurs for
moderate and large $t/J$ at small $\delta$ and has
$\Delta_x = - \Delta_y$ and has been considered previously
by
Kotliar and Liu~\cite{kliu}. The quasi-particle gap vanishes
at 4
isolated points
$(\pm k^0 , \pm k^0 )$ in the Brilluoin zone.
\newline
({\it iv\/}) $s^{\ast}$-wave superconductor:
Finally for large $\delta$ we obtain an extended $s$-wave solution
with no on-site pairing which has
$\Delta_x = \Delta_y$. There is again a fully gapped
excitation
spectrum; for large $t\delta/J$,
the gap vanishes as $t\delta \exp ( - ct\delta / J )$ where
$c$ is
a constant of order unity.

Finally we address the issue of the stability of the phases
discussed
above towards phase separation.
A realistic model of the high-$T_c$ materials should also
include
the long-range Coulomb interactions between the electrons
and
between the rare-earth ions and the electrons. These
interactions
will reduce any tendency towards phase separation. It is
nevertheless
interesting to examine this issue for $H_{tJ}$ alone. The
results of a Maxwell construction on the
energy-versus-doping
curve are shown in Fig~\ref{fig5}. The region below the dotted line
is susceptible to separation into an insulating
antiferromagnet
with $\delta = 0$ and a hole-rich phase with $\delta$ on the
dotted
line. The boundary towards phase separation approaches
$\delta = 0 $
as $t/J \rightarrow \infty$, in agreement with
Ref~\cite{emery}.
However at small values of $t/J$, in particular at $t=0$,
we find no phase separation,
unlike Emery {\em et. al.}~\cite{emery}. The absence of
phase
separation at $t=0$ in the present calculation
is due to the extra stability acquired at large $N$
by fully dimerized
solutions with every site less than half-filled.

\section*{\Large\bf 5. Conclusions}

This paper has presented details on the application of a new
large $N$ expansion based on the symplectic groups $Sp(N)$
which was introduced recently~\cite{self3}. These groups
naturally generalize to all values of $N$ the physics of
$SU(2) \cong Sp(1)$ antiferromagnetism: the pairing of spins
with opposite directions in spin space. They lead to a
particularly appealing treatment of frustrated
antiferromagnets with all antiferromagnetic exchange
interactions treated democratically. The large $N$ limit
was taken by placing spins on each site which transform as
$n_b$ symmetric products on the fundamental of $Sp(N)$, and
fixing $n_b / N$ at an arbitrary value ($n_b = 2S$ for
$SU(2)$). For large $n_b / N$ the ground states become
identical to the classical solutions and possess magnetic
long range order (LRO). Upon reducing $n_b / N$ all the LRO
states under continuous transition to short range ordered
(SRO) states whose properties are closely connected to the
LRO states. In particular for the square lattice model with
first, second, and third neighbor interactions (Figs~\ref{fig1}, \ref{fig2},
\ref{fig3}, \ref{fig4})
we found two
distinct classes of states:
\newline
{\em (i)} LRO states with commensurate, collinear order
led at smaller values of $n_b / N$ to SRO states with a gap
to all excitations, spinons confined by a compact $U(1)$
gauge force, and spin-Peierls or valence-bond-solid order
controlled by the value of
$n_b \mf$.
\newline
{\em (ii)} LRO states with incommensurate, coplanar order
led at smaller values of $n_b / N$ to SRO states with a gap
to all excitations, unconfined spinons and no spin-Peierls
order.
A charge $-2$
scalar in a Higgs phase was responsible for liberating the
spinons~\cite{fradkin_shenker}.
These incommensurate SRO states appear to contradict the
`fractional quantization principle' of Laughlin~\cite{laugh}.
This principle asserts that spin-1/2 excitations of
a non-degenerate `spin-liquid' ground state must
interact with a gauge force which endows them
with fractional statistics. The SRO states found in this paper
have a two-fold degeneracy but this can easily be lifted
by adding small explicit symmetry breaking terms to $H_{AF}$
(this can be done for the $(q,\pi)$-SRO phase by
modifying $J_3$ on just the $x$-directed links, and for
the $(q , q )$-SRO phase by changing $J_2$ on the links
in the $(1,1)$ direction).
The  modified model now has a non-degenerate ground state,
no gapless excitations, and unconfined spin-1/2 spinon
excitations with bosonic statistics,
contradicting Ref~\cite{laugh}.
\newline
Additional intermediate phases between the Higgs and
confinement SRO phases are possible and are presently under
investigation. No chirally ordered~\cite{wwz} or
spin-nematic~\cite{premi} states were found.

An initial discussion of the application of the symplectic
groups to doped antiferromagnets was also presented. In this
case they allow a natural implementation of both
antiferromagnetic exchange and hole hopping terms, and
emphasize the natural connection between antiferromagnetism
and singlet superconductive pairing~\cite{anderson}. In
particular a study of the $T=0$, $N=\infty$ limit of a $t-J$
model with spins represented by fermions and holes by
spinless bosons yields superconducting ground states (Fig~\ref{fig5})
in much
the same manner as Ref~\cite{bza}.
Extension of these results to finite temperature and finite
$N$ is clearly of great interest and is currently in
progress.

\section*{\large\bf Acknowledgments}

We thank D.P. Arovas, P. Chandra, P. Coleman, E. Dagotto, G. Kotliar,
J.B. Marston, A. Moreo, R. Shankar, and Z. Wang
for useful
discussions. This work was supported by the A. P. Sloan
Foundation
(N.R. and S.S.) and by NSF PYI Grant DMR 8857228 (S.S.).

\newpage
\end{document}